\documentclass[%
 reprint,
 superscriptaddress,
 amsmath,amssymb,
 aps,
 pra,
 longbibliography,
]{revtex4-2}

\usepackage{graphicx}
\usepackage{dcolumn}
\usepackage{bm}
\usepackage{mathtools}
\usepackage{hyperref}
\usepackage{xcolor}
\usepackage{listings}
\usepackage{bbold}
\usepackage{amsthm}
\usepackage{physics}
\usepackage{bibunits}

\defaultbibliography{apssamp}
\defaultbibliographystyle{apsrev4-2}

\newcommand{\subfigref}[2][]{\hyperref[#2]{\ref*{#2}(#1)}}
\newcommand{\transition}[2]{\(#1\text{--}#2\)}

\makeatletter
\newcommand*{\balancecolsandclearpage}{%
  \close@column@grid
  \clearpage
  \twocolumngrid
}
\makeatother
\begin{document}

\preprint{APS/123-QED}

\begin{bibunit}
\title{Carrier Revival in Long Trapped-Ion Chains}

\author{Florian Egli}
\affiliation{Max-Planck-Institut für Quantenoptik, Hans-Kopfermann-Straße 1, 85748 Garching, Germany}
\author{Chris Shanks}
\affiliation{Department of Physics, Texas A\&M University, College Station, Texas 77843-4242, USA}
\author{James Bounds}
\affiliation{Department of Physics, Texas A\&M University, College Station, Texas 77843-4242, USA}
\author{Jorge Moreno}
\affiliation{Max-Planck-Institut für Quantenoptik, Hans-Kopfermann-Straße 1, 85748 Garching, Germany}
\author{Muhammad Thariq}
\affiliation{Max-Planck-Institut für Quantenoptik, Hans-Kopfermann-Straße 1, 85748 Garching, Germany}
\author{Erdem Yilmaz}
\affiliation{Max-Planck-Institut für Quantenoptik, Hans-Kopfermann-Straße 1, 85748 Garching, Germany}

\author{Theodor W. Hänsch}
\affiliation{Max-Planck-Institut für Quantenoptik, Hans-Kopfermann-Straße 1, 85748 Garching, Germany}
\affiliation{Fakultät für Physik, Ludwig-Maximilians-Universität München, Schellingstraße 4, 80799 Munich, Germany}

\author{Thomas Udem}
\affiliation{Max-Planck-Institut für Quantenoptik, Hans-Kopfermann-Straße 1, 85748 Garching, Germany}
\affiliation{Fakultät für Physik, Ludwig-Maximilians-Universität München, Schellingstraße 4, 80799 Munich, Germany}

\author{Akira Ozawa}
\affiliation{Institute for Multidisciplinary Sciences, Yokohama National University, Yokohama 240-8501, Japan}
\affiliation{Max-Planck-Institut für Quantenoptik, Hans-Kopfermann-Straße 1, 85748 Garching, Germany}

\date{\today}

\begin{abstract}
For a single trapped ion, the excitation spectrum of a narrow optical transition consists of a Doppler‑ and recoil‑free carrier accompanied by motional sidebands, which are equally spaced by the trap secular frequency and lie under a Doppler‑broadened envelope that is shifted by the photon recoil. Outside the Lamb--Dicke regime, the large photon recoil distributes the line strength across many sidebands and suppresses excitation of the carrier. With multiple ions, the motional spectrum becomes dense, and the carrier is further weakened.
Here, we predict a counterintuitive revival effect: increasing the number of ions in a linear chain can restore strong carrier excitation even under trapping conditions far from the single-ion Lamb--Dicke regime. Using a quantum‑mechanical model of the excitation dynamics in linear ion chains, we find that sufficiently long chains concentrate the spectrum into the carrier. This effect enables efficient excitation of light ions at short wavelengths. It may also benefit multi‑ion optical clocks and mixed‑species quantum‑logic spectroscopy.

\end{abstract}

\maketitle

Ions confined and cooled in the isolated environment of a radio-frequency Paul trap provide an ideal platform for various applications, including precision spectroscopy for probing fundamental physics and optical frequency standards \cite{ludlow_optical_2015,moreno_toward_2023,kozlov_highly_2018}.
In the quantum picture, the laser excitation of the internal electronic state of a single ion is coupled to its motion by the recoil of the photon. In a quantum harmonic oscillator, the energy eigenstates have energies $E_n = \hbar \omega_\mathrm{sec}(n+1/2)$. The excitation spectrum of a single trapped ion is therefore composed of a carrier that is free of the Doppler and recoil shift, and motional sidebands separated by the trap secular frequency $\omega_{\mathrm{sec}}$
\cite{note1}. Detuning the laser from the carrier by integer multiples of the secular frequency $\delta=\Delta n\,\omega_\mathrm{sec}$ allows exciting the electronic transition while also changing the occupation number of the harmonic oscillator from $n$ to $n+\Delta n$.
The sidebands appear under a Doppler envelope with the width---just as with free particles---given by the temperature $T$, or more precisely, for single ions, by the motional energy.
The center of the Doppler envelope is shifted by the recoil frequency $\omega_\mathrm{rec}=\hbar k^2/(2m)$ of the excitation, where $m$ is the ion mass and $k=2\pi/\lambda$ is the wave vector of the excitation laser. This is illustrated in Fig.~\subfigref[a]{fig:spectrum_panel}.

The line strengths of the sidebands sum up to the line strength of a rigidly fixed ion, whose excitation spectrum only consists of a single carrier transition \cite{herrmannFeasibility2009, supplement_arxiv}.
Therefore, when a narrow-linewidth continuous-wave laser is used to excite either the carrier or a selected sideband of a trapped ion, the strength of that transition is reduced.
Under the condition that the Doppler envelope is narrower than the trap secular frequency
the excitation spectrum is dominated by a single spectral component that recovers the full line strength. This condition defines the Lamb--Dicke regime \cite{leibfried_quantum_2003,wineland_experimental_1998}, and can also be stated as the regime in which the extension of the ion's wavefunction is much smaller than the laser wavelength.

For a single ion in a harmonic potential, the coupling of the laser-atom interaction to the motion is described by the interaction Hamiltonian $\hat{H}_\mathrm{int}\propto \exp[ik\hat{x}]=\exp[ikx_0(\hat{a}+\hat{a}^\dag)]$, where $\hat{a}$ and $\hat{a}^\dag$ are the annihilation and creation operators of the harmonic oscillator and $x_0=\sqrt{\hbar/(2m\omega_\mathrm{sec})}$ is the extension of the ground state wavefunction. Introducing the Lamb--Dicke parameter $\eta=kx_0=\sqrt{\omega_\mathrm{rec}/\omega_\mathrm{sec}}$, the condition for the Lamb--Dicke regime is $\eta\sqrt{\langle(\hat{a}+\hat{a}^\dag)^2\rangle}=\eta\sqrt{2\bar{n}+1}\ll1$, where $\bar{n}=\langle \hat{a}^\dag \hat{a}\rangle$ is the expectation value of the number operator. In the Lamb--Dicke regime $H_\mathrm{int}\propto1+i\eta(\hat{a}+\hat{a}^\dag)$, resulting in a strong carrier transition accompanied by weaker red and blue sidebands.
The carrier transition is usually required for precision spectroscopy, as its frequency is independent of the secular motion and therefore does not require precise knowledge or control of the trap voltages and dimensions.

\begin{figure*}[!t]
\includegraphics[width=\textwidth]{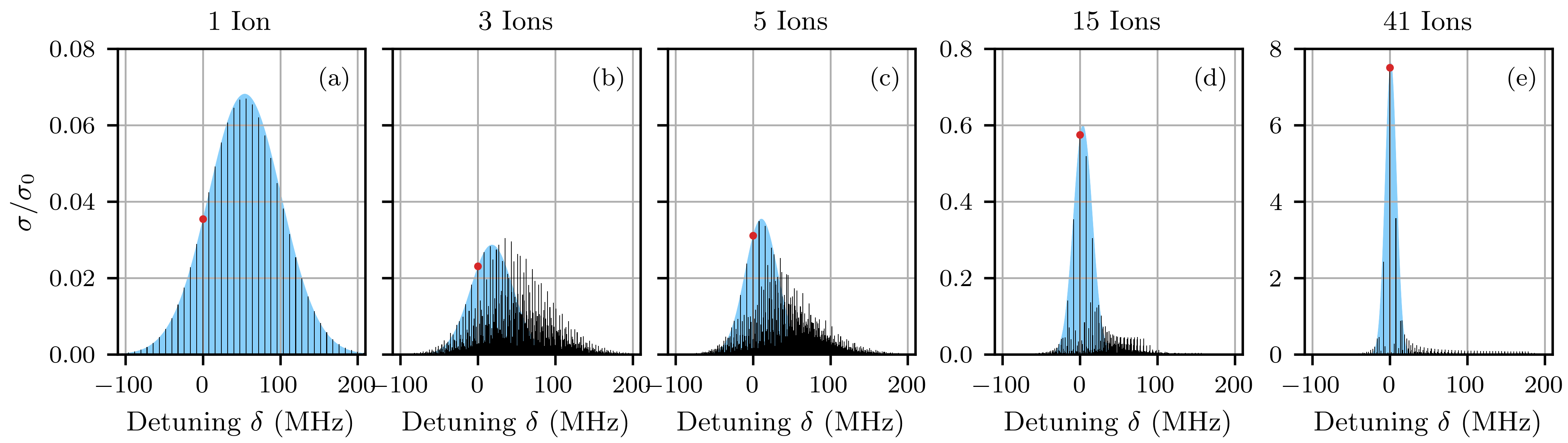}
\caption{\label{fig:spectrum_panel} The calculated excitation spectra of the \transition{1S}{2S} two-photon transition for He$^+$ ions driven by two 60.8~nm photons co-propagating along the linear chain, as an example for the carrier revival effect,
according to \eqref{eq:total_cross_section}. The number of ions $N$ are 1(a), 3(b), 5(c), 15(d) and 41(e). We choose an axial secular frequency $\omega_\mathrm{sec}$ of $2\pi\times 8~\text{MHz}$ and set the chain's temperature to $T=1~\text{mK}$.
The recoil frequency is $\omega_\mathrm{rec}=2\pi\times54\,\mathrm{MHz}$, resulting in a single-ion Lamb--Dicke parameter of $\eta=2.6$. This example is illustrative: The required radial confinement to maintain a linear chain of $N=41$ ions is $\omega_\mathrm{r}\gtrsim 0.715\,N^{0.838}\,\omega_\mathrm{sec}=2\pi\times128\,\mathrm{MHz}$  \cite{wineland_experimental_1998, jamesQuantum1998}, which exceeds current experimental capabilities. The red dot marks the height of the carrier transition. The vertical axis is the line strength in units of the cross section $\sigma_0$ of a single, rigidly fixed ion. The plots (a-c) have the same scale, while the vertical axes of the plots (d) and (e) are scaled by a factor 10 and 100, respectively. For modest ion numbers of 3(b) and 5(c), the sideband spectrum becomes more complicated, and the strength of many individual sideband decreases. However, with more ions, the sideband structure becomes cleaner again and the carrier strength revives and grows with the number of ions in the chain.
The filled blue curve is a Gaussian envelope with its width given by the Doppler broadening and shifted to the blue by the recoil frequency $\omega_\mathrm{rec}$, with an effective mass given by $m_\text{eff}=Nm$. It is scaled such that the cross section of the carrier transition lies on the curve.
}
\end{figure*}
For many applications, such as precision spectroscopy, optical frequency standards, and coherent control for quantum computation, it is desirable to increase the number of atoms/ions in order to improve statistics and scalability.
For optical frequency standards using neutral atoms, this was enabled by the invention of the lattice clock \cite{takamoto_optical_2005}, which combines low systematics with good statistics.
For ion clocks, the concept of a multi-ion clock was proposed to enable highly accurate clocks with low statistical uncertainty \cite{herschbach_linear_2012}. It is therefore particularly important to investigate the excitation spectrum of trapped many-ion systems.

In a linear ion trap, a chain of $N$ ions has $N$ axial vibrational modes. The excitation spectrum comprises all possible combinations of occupation number changes in these normal modes. With trap parameters, where a single ion is far outside the Lamb--Dicke regime, this leads to a strong increase in the density of motional sidebands as more ions are added, as shown in Fig~\subfigref[b]{fig:spectrum_panel}.

This issue is particularly important for transitions at short wavelengths, where achieving the Lamb--Dicke regime is more challenging. This is the case, for example, in our planned spectroscopy experiment of the \transition{1S}{2S} two‑photon transition in $\mathrm{He}^+$ at 60.8~nm, when driven with co‑propagating photons \cite{moreno_toward_2023, herrmannFeasibility2009}. Another example is the nuclear transition in trapped $^{229}\mathrm{Th}$ ions, which can be driven with a short wavelength laser \cite{campbell_single-ion_2012, tiedau_laser_2024}.

We show that, using a large enough number of ions, the carrier strength revives and further increases with the ion number until the excitation spectrum eventually becomes dominated by a single carrier transition, see Fig.~\subfigref[c-e]{fig:spectrum_panel}.
Intuitively, the recoil momentum imparted to a single ion in a long ion chain becomes inefficient at driving quantized collective motion, especially for higher-order modes. This behavior is closely analogous to the Mössbauer effect.

We investigate this effect theoretically and give an analytic expression that can be used to efficiently compute the line strength of any sideband, including the carrier.
We assume a linear ion trap where the ions are strongly confined in the radial directions but weakly confined by a harmonic potential along the $x$ direction. When the radial confinement is sufficiently strong, the ions form a linear chain. The potential energy of an ion chain with $N$ identical ions of mass $m$ and charge $Ze$ in the trap is given by
\begin{equation}
V = \sum_{i = 1}^{N}\frac{1}{2}m \omega_\text{sec}^2 x_{i}(t)^2 + \sum_{\substack{i,j=1 \\ {i\neq j}}}^N\frac{(Ze)^2}{8\pi\epsilon_0}\frac{1}{|x_{i}(t)-x_{j}(t)|}\, ,
\end{equation}
where $x_i$ is the axial position of the $i$-th ion. From the potential energy, classical equations of motion yield the axial eigenfrequencies $\omega_\alpha$ and the complex amplitudes $\beta^\alpha_i$ of the $i$-th ion for the $\alpha$-th axial mode \cite{jamesQuantum1998}. We assume that the laser is aligned parallel to the ion chain. We can define the generalized Lamb--Dicke parameters \cite{morigiTwospecies2001, herrmannFeasibility2009}
\begin{equation}
\label{eq:gen_lamb_dicke}
\eta_i^\alpha = k\,\beta_i^\alpha\sqrt{\frac{\hbar}{2m\omega_\alpha}}\, ,
\end{equation}
which characterize the strength of the coupling of the excitation of the $i$-th ion to the motion of mode $\alpha$. Note that for transitions where multiple photons are absorbed, $k$ should be taken as the magnitude of the vector sum of the wavevectors of all involved excitation photons. A laser beam addressing the $i$-th ion can excite all modes for which $\eta_i^{\alpha}$ is nonzero.

Using the generalized Lamb--Dicke parameters, the interaction Hamiltonian $\hat{H}_{\mathrm{int}}(t)$ is readily extended to the multi-ion case. The transition matrix element $M_i$ between the initial motional state
$\ket{\mathbf{n}} = \ket{n_1,n_2, ..., n_N}$ and the final state $\ket{\mathbf{n}+\Delta\mathbf{n}} = \ket{n_1+\Delta n_1,n_2+\Delta n_2, ..., n_N+\Delta n_N}$ is given by a product of single-mode matrix elements as
\cite{herrmannFeasibility2009, morigiTwospecies2001}
\begin{equation}
\label{eq:MatriexElement}
M_i(\mathbf{n}, \Delta\mathbf{n})=\prod_{\alpha=1}^N\bra{n_\alpha+\Delta n_\alpha} e^{i\eta_i^\alpha (\hat{a}_\alpha+\hat{a}_\alpha^\dagger)}\ket{n_\alpha}.
\end{equation}
Note that the matrix element $M_i$ is dependent on both the initial state $\ket{\mathbf{n}}$ and the occupation-number changes $\Delta\mathbf{n}= (\Delta n_1, \Delta n_2, \ldots, \Delta n_N)$.
We assume dipole-dipole interactions \cite{vogtCollective1996, devoeObservation1996}, coupling the internal states of the ions, can be ignored. This assumption is valid in most ion-trap setups \cite{morigiLaser1999}.
To quantify the line strength of a sideband $\Delta\mathbf{n}$, we compute the cross section, averaged over the initial state distribution $P(\mathbf{n})$
\begin{equation}
\label{eq:cross_section}
\sigma_i(\Delta \mathbf{n})=\sigma_0 \sum_{\mathbf{n}} \abs{M_i(\mathbf{n},\Delta\mathbf{n})}^2 P(\mathbf{n})\, ,
\end{equation}
where the sum includes all allowed initial states $\ket{\mathbf{n}}$, such that \(n_\alpha+\Delta n_\alpha\geq 0\) for all $\alpha$. The cross-section of a single, rigidly fixed ion is given by $\sigma_0$.
Further, we assume $P(\mathbf{n})$ to be the thermal distribution, given by
\begin{equation}
    P(\mathbf{n}) = \prod_{\alpha=1}^N\left(1-\exp[-\frac{\hbar\,\omega_\alpha}{k_\text{B} T}]\right)\exp[-n_\alpha \frac{\hbar\,\omega_\alpha}{k_\text{B} T}]\, .
\end{equation}
With this, we derive a closed form for the cross section \cite{supplement_arxiv}
\begin{equation}
\label{eq:cross_section_analytic}
\begin{aligned}
\sigma_i(\Delta\mathbf{n})&=\sigma_0 \prod_{\alpha=1}^N e^{-\left(\eta_i^\alpha\right)^2\,(1+2\bar{n}_\alpha)}\left(\frac{\bar{n}_\alpha}{\bar{n}_\alpha+1}\right)^{-\Delta n_\alpha/2}\\
&\qquad\qquad\times I_{\Delta n_\alpha}\left(2\left(\eta_i^\alpha\right)^2\sqrt{\bar{n}_\alpha(\bar{n}_\alpha+1)}\right)\, ,
\end{aligned}
\end{equation}
in close analogy to the derivation of the single-ion cross section given in \cite{winelandLaser1979}. Here, $I_{\Delta n_\alpha}(z)$ is the modified Bessel function of the first kind and order $\Delta n_\alpha$, and $\bar{n}_\alpha=(\exp[\hbar\omega_\alpha/(k_\mathrm{B}T)]-1)^{-1}$ is the mean thermal occupation number of mode $\alpha$.
Using \eqref{eq:cross_section_analytic}, one can readily calculate the cross section of any sideband $\Delta\mathbf{n}$. From the vector of occupation number changes $\Delta\mathbf{n}$ the detuning $\delta=\sum_{\alpha=1}^N\Delta n_\alpha \omega_\alpha$ can be dermined. Since the eigenfrequencies $\omega_\alpha$ are incommensurate \cite{morigiLaser1999}, there are no two sidebands $\Delta\mathbf{n}$ that share the same $\delta$.

We consider a laser beam that is aligned to the axis of the ion chain and addresses all ions equally. We define the total cross section of the ion chain as the incoherent sum of all single ion contributions
\begin{equation}
\label{eq:total_cross_section}
\sigma(\Delta \mathbf{n})=\sum_{i=1}^N \sigma_i(\Delta \mathbf{n})\, .
\end{equation}
This is a good approximation for example in spectroscopy experiments with a decoherence process that is faster than the Rabi dynamics: then, simultaneous excitation events of multiple ions can be ignored. In this description we assume that the probe laser does not modify the vibrational state of the ion crystal, and no Rabi flopping or time evolution is considered. In the weak--excitation limit, \(\sigma(\Delta \mathbf{n})\) corresponds directly to the excitation-probability spectrum.

As an example, we investigate the \transition{1S}{2S} transition in trapped \(\mathrm{He}^+\) ions, using two co-propagating photons at \(60.8~\text{nm}\).
The magnitude of the effective wavevector is given by $k=2\times 2\pi/(60.8~\text{nm})$ and we assume an axial secular frequency of \(\omega_{\mathrm{sec}} = 2\pi \times 8~\text{MHz}\) and an ion-chain temperature of \(T=1~\text{mK}\), which is the temperature of the He$^+$ ions in our setup.
The resulting single-ion Lamb--Dicke parameter is $\eta=2.6$, far from the Lamb--Dicke regime.
The excitation spectra according to \eqref{eq:total_cross_section} for sidebands larger than $10^{-6}\sigma_0$ are shown in Fig.~\ref{fig:spectrum_panel}.

For a single ion (see~Fig.~\subfigref[a]{fig:spectrum_panel}), the spectrum consists of many modes separated by the axial secular frequency $\omega_\mathrm{sec}$. The sideband spectrum nicely fits a Gaussian envelope (shaded blue curve) with its FWHM given by the Doppler broadening $\Delta \omega=k\sqrt{8k_\text{B} T\ln(2)/m}\approx2\pi\times112~\text{MHz}$ and shifted to the blue by the recoil frequency $\omega_\mathrm{rec}\approx 2\pi\times54.0~\text{MHz}$. Because of the recoil shift, the carrier---which corresponds to the transition with $\Delta \mathbf{n}=(0,0,\ldots,0)\equiv\mathbf{0}$---is not the strongest line. 
Note that the axial secular frequency $\omega_{\mathrm{sec}}$ determines the spacing between the sidebands, however, it does not affect either the width or the shift of the Doppler broadening envelope.

Adding more ions significantly complicates the spectrum of the system. The number of sidebands with non-negligible strength increases very rapidly and the excitation strength of each individual sideband reduces correspondingly. The spectrum becomes much more densely populated and the carrier transition is suppressed as the number of ions increases. However, this trend does not persist for large number of ions. 
Beyond a certain ion number, the carrier strength revives, suppressing the sideband structure. For sufficiently large ion numbers the spectrum becomes dominated by the carrier, with only a few residual sidebands remaining in its vicinity. For $N=41$ ions, the total carrier strength reaches $\sigma(\mathbf{0})\gtrsim7.5\sigma_0$.
This reflects a roughly 200-fold enhancement of the carrier strength compared to a single ion ($0.036 \sigma_0$), as shown in Fig.~\subfigref[a]{fig:spectrum_panel}. The enhancement is even more substantial when compared to the three-ion case, where the carrier cross section is suppressed to $0.023\sigma_0$.
This non-monotonic behavior is the result of two competing processes: the exponential increase of the number of sidebands on the one hand, and the reduction of the generalized Lamb--Dicke parameters on the other. The latter can be seen from Eq.~\eqref{eq:gen_lamb_dicke}: The generalized Lamb--Dicke parameters are proportional to the amplitudes of the normal modes $\beta_i^\alpha$, which approximately scale as $\beta_i^\alpha \sim 1/\sqrt{N}$ \cite{morigiLaser1999, jamesQuantum1998}. Intuitively, the reduction of the generalized Lamb--Dicke parameters reflects the effective inertia that grows as the chain becomes longer.
The recoil of an absorbed photon is shared among all ions in the chain, making the system more rigid against motional excitation.
This is visualized in Fig.~\ref{fig:spectrum_panel} by the shaded blue curve, which corresponds to the Doppler-broadened spectrum of a fictitious free particle with effective mass $m_\text{eff} = N m$ and temperature $T=1\,\mathrm{mK}$. It shows good agreement with the spectrum for both single ions and large ion numbers, as the center-of-mass mode dominates. However, in the intermediate regime shown in Fig.~\subfigref[b,~c]{fig:spectrum_panel}, where additional motional modes contribute significantly and the spectrum becomes denser, the model fails to provide an accurate description.

Another important contribution to the carrier revival effect is that the new axial modes introduced by adding more ions have eigenfrequencies $\omega_\alpha$ that monotonically increase with $N$. These high-frequency modes therefore have smaller Lamb--Dicke parameters than the lower-frequency eigenmodes, making motional excitations involving them less pronounced. In addition, under a thermal distribution of initial populations, these high-frequency modes are less populated at a given temperature, further reducing their contributions to the sidebands.

Calculating the full spectrum is computationally expensive.
Even when limiting occupation-number changes to $\Delta n_\alpha = -s_{\min}, \ldots, s_{\max}$ per mode, the resulting spectrum consists of all combinations across $N$ modes and thus contains a total of $(s_{\min} + s_{\max} + 1)^N$ sidebands, and hence grows exponentially with $N$. Because many sidebands are very weak and contribute only marginally to the spectrum, we evaluate only those whose contributions exceed a cutoff $\epsilon$, i.e., those with $\sigma_i(\Delta\mathbf{n})/\sigma_0 > \epsilon$.
A computationally efficient algorithm for selecting these lines is described in the Supplemental Material \cite{supplement_arxiv}. The algorithm reduces the number of sidebands that need to be calculated drastically, making the computational evaluation of the spectrum possible: In the case of the 15 ion spectrum shown in Fig.~\subfigref[c]{fig:spectrum_panel}, the occupation number change per mode can be limited to $\Delta n_\alpha=-7,\ldots,9$ and the naive number of sidebands to be calculated would be $17^{15}=2.86\times10^{18}$. This can be reduced to $1.2\times10^5$ with our algorithm, when we set the cutoff to $\epsilon=10^{-6}$. The sum across all sidebands is a good metric for the fidelity of the approximation \cite{herrmannFeasibility2009, supplement_arxiv}. In the evaluation of Fig.~\ref{fig:spectrum_panel}, the neglected sidebands contribute less than $5\%$ of the total line strength.

\begin{figure}[ht!]
\includegraphics[width=\columnwidth]{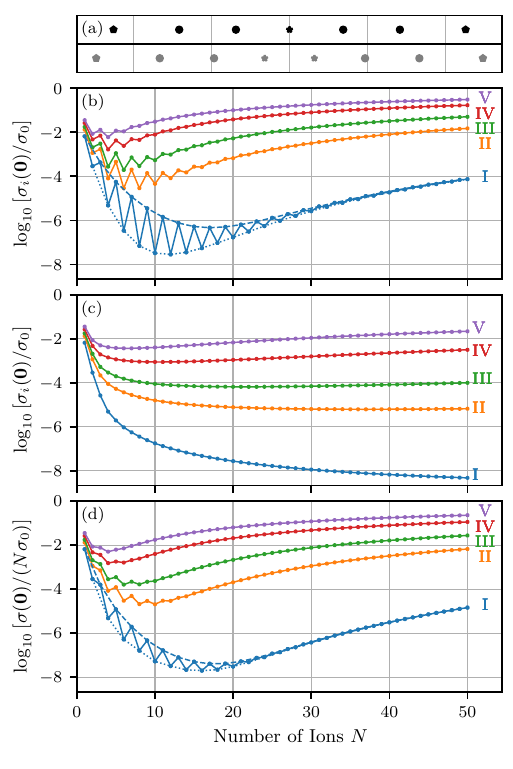}
\caption{\label{fig:carrier_panel} The cross section of the carrier, defined by $\Delta\mathbf{n}=\mathbf{0}$, normalized to the cross-section $\sigma_0$ of a single, rigidly fixed ion, as a function of the number of ions in the chain. In (a), ion chains are shown with an odd number ($N=7$) of ions in black at the top and with an even number ($N=8$) of ions in gray on the bottom. The center ions are marked by stars and the end ions by pentagons. The carrier is evaluated for (b) the center ion with the index $i=\lceil N/2\rceil$, (c) the end ion with the index $i=1$ or $i=N$, and (d) the average over all ions, for different axial secular frequencies $\omega_\text{sec}$: (I) $2\pi\times1.5~\text{MHz}$, (II) $2\pi\times3~\text{MHz}$, (III) $2\pi\times4~\text{MHz}$, (IV) $2\pi\times6~\text{MHz}$, (V) $2\pi\times8~\text{MHz}$. The temperature and the recoil are the same as used in Fig.~\ref{fig:spectrum_panel}. The resulting single-ion Lamb--Dicke parameters $\eta$ are: (I) $6.0$, (II) $4.2$, (III) $3.7$, (IV) $3.0$, and (V) $2.6$. After an $\omega_\text{sec}$-dependent turnaround point, the carrier for all three cases revives and starts growing again, and eventually exceeds the single ion value. For the end ion, the turnaround point for $\omega_\text{sec}=2\pi\times1.5\,\text{MHz}$ lies approximately at 150 ions (not shown).}
\end{figure}

To study the evolution of the spectrum---from many weak lines to a few strong sidebands, including the carrier---we examine the carrier strength of individual ions in the chain, characterized by $\sigma_i(\mathbf{0})$ calculated with Eq.~\eqref{eq:cross_section_analytic}, as a function of the number of ions $N$.
The carrier strength for the central ion, the ion at the end of the chain, and the average over all ions is shown in Fig.~\ref{fig:carrier_panel}. The center ion is defined as the one closest to the geometrical center of the ion chain. For all ions, the carrier first diminishes as $N$ increases and then undergoes carrier revival after reaching a turnaround point dependent on ion position and $\omega_\mathrm{sec}$. The center ion shows the strongest carrier revival effect, with the earliest onset.
Its carrier strength is stronger for odd ion numbers than for even ion numbers, see Fig.~\subfigref[b]{fig:carrier_panel}.
This is because for odd ion chains a single ion sits exactly at the trap center and does not participate in modes that are antisymmetric about that center (e.g., the breathing mode), so its motional coupling is suppressed. For even chains the two central ions are displaced symmetrically from the center and participate in all motional modes, reducing the carrier strength relative to odd ion numbers.

The ion number at which the carrier strength starts to revive is set by the balance between the exponentially growing number of sidebands and their suppression due to the reduced Lamb--Dicke parameter. The turnaround ion number is empirically determined by $\eta\sqrt{k_\text{B} T /(2\hbar \omega_\mathrm{sec})}\sim \sqrt{\omega_\mathrm{rec}}/\omega_\mathrm{sec}$, where $\eta$ is the single-ion Lamb--Dicke parameter. Note that the turnaround point is independent of the ion mass $m$. This expression applies well to the case of the cross-section of the average ion, for even $N$ (e.g., the dotted line in Fig.~\subfigref[d]{fig:carrier_panel}). A comparison of the empirical expression and the turnaround point extracted from the computational evaluation is shown in Fig.~\ref{fig:turnaround_point}.
For the given parameter space, the difference between the empirical expression and the extracted turnaround point is less than 1.5 ions. Therefore, the formula can be used to approximately evaluate trap setups for possible carrier revival regimes.

\begin{figure}[ht!]
\includegraphics[width=\columnwidth]{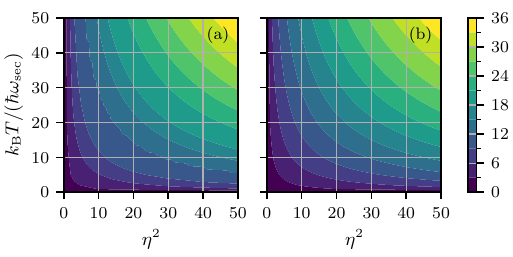}
\caption{\label{fig:turnaround_point} The turnaround ion number for the carrier of the average ion, when considering only even ion numbers, as a function of $\eta^2$ and $k_\text{B} T/(\hbar\omega_\text{sec})$, (a) extracted from the computational evaluation via finding the minimum of a cubic spline to the data (see dotted line in Fig.~\subfigref[d]{fig:carrier_panel}), and (b) from the empirical formula $\eta\sqrt{k_\mathrm{B}T/(2\hbar\omega_\mathrm{sec})}$. The absolute difference between the formula and the minimum of the cubic spline is less than $1.5$ ions for the shown parameter range.}
\end{figure}

In this work, we have shown that it is possible to reach an effective Lamb--Dicke regime using axially addressed, long linear ion chains, even with trap parameters where a single trapped ion has a Lamb--Dicke parameter $\eta > 1$.
For sufficiently long ion chains, the spectrum becomes dominated by only three peaks: the carrier, as well as the first red and blue sidebands of the center-of-mass mode, which is the lowest-energy eigenmode ($\omega_1 = \omega_\text{sec}$), as shown in Fig.~\subfigref[e]{fig:spectrum_panel} for $N=41$. In \cite{morigiLaser1999}, it was discussed that the Lamb--Dicke regime can be reached for each individual mode at large ion numbers $N$, and that these modes can be considered decoupled for cooling applications. Our results go even further: for sufficiently large $N$, it is possible to reach an effective Lamb--Dicke regime for the entire ion chain
which suppresses sidebands from all but the lowest-frequency mode. The resulting spectrum is thus analogous to the single-ion Lamb--Dicke regime, with the carrier and only two weaker sidebands needing to be taken into account.

This approach opens new possibilities for reaching the Lamb--Dicke regime and offers a wide range of potential applications. Precision spectroscopy in the VUV and XUV regimes, particularly for light ions, requires very tight confinement to achieve the Lamb--Dicke regime. The present scheme enables strong carrier addressing by employing sufficiently long ion chains. Examples of such systems include the \transition{1S}{2S} spectroscopy of He$^+$ ions and excitation of the VUV nuclear transition of $^{229}$Th ions. Multi-ion optical frequency standards \cite{keller_controlling_2019, hauserClock2025} could also benefit from this scheme, as it allows strong carrier transitions that scale with the number of trapped ions while working with modest secular frequencies.
This application requires extending our model to the Rabi dynamics of
$N$ simultaneously driven ions, including the time evolution of the motional states. Preliminary results of our theoretical investigation suggest the carrier revival effect persists in these systems.

This effect is also expected in ion-trap configurations that go beyond simple linear chains composed of a single ion species. Our numerical study shows that linear chains containing multiple ion species also exhibit carrier revival behavior, which could find applications in efficient quantum logic spectroscopy \cite{schmidt_spectroscopy_2005}. In a typical sympathetic cooling geometry, spectroscopy ions may form a linear chain surrounded by heavier coolant ions. The axial motion of these spectroscopy ions is expected to be only weakly coupled to the motion of the coolant ions. Consequently, carrier revival is still expected for axial excitation of the spectroscopy ions. Carrier revival may also remain effective for radial excitations of sympathetically cooled ions, as the radial motion experiences increased inertia when a sufficiently large number of coolant ions is used, resulting in reduced effective Lamb--Dicke parameters for the radial modes.
The recently demonstrated measurement of sympathetically cooled molecular ions could, in fact, already be influenced by this effect \cite{kortunovProton2021}.
\nocite{batemanHigher2023}

\begin{acknowledgments}
We express our deep gratitude to Hans Schuessler, who made significant contributions to our ion-trapping projects.
We would like to thank Tanja Mehlstäubler and Dietrich Leibfried for constructive feedback on the manuscript.
This project has received funding from the European Research Council (ERC) under the European Union’s Horizon 2020 research and innovation programme (Grant Agreement No. 742247)
and the Qlock project (“High-Precision Laser Systems for Quantum Control and Quantum Clocks”), which is part of Munich Quantum Valley.
This work was funded by the Deutsche Forschungsgemeinschaft (DFG, German Research Foundation) – project number 535639992.
T.~W.~Hänsch acknowledges support from the Max Planck Foundation.
A.~Ozawa acknowledges support from JSPS KAKENHI (Grant No. 24K23943).  
\end{acknowledgments}
\putbib
\balancecolsandclearpage

\end{bibunit}

\begin{bibunit}

\title{Supplement to Carrier Revival in Long Trapped-Ion Chains}
\author{Florian Egli}
\affiliation{Max-Planck-Institut für Quantenoptik, Hans-Kopfermann-Straße 1, 85748 Garching, Germany}
\author{Chris Shanks}
\affiliation{Department of Physics, Texas A\&M University, College Station, Texas 77843-4242, USA}
\author{James Bounds}
\affiliation{Department of Physics, Texas A\&M University, College Station, Texas 77843-4242, USA}
\author{Jorge Moreno}
\affiliation{Max-Planck-Institut für Quantenoptik, Hans-Kopfermann-Straße 1, 85748 Garching, Germany}
\author{Muhammad Thariq}
\affiliation{Max-Planck-Institut für Quantenoptik, Hans-Kopfermann-Straße 1, 85748 Garching, Germany}
\author{Erdem Yilmaz}
\affiliation{Max-Planck-Institut für Quantenoptik, Hans-Kopfermann-Straße 1, 85748 Garching, Germany}

\author{Theodor W. Hänsch}
\affiliation{Max-Planck-Institut für Quantenoptik, Hans-Kopfermann-Straße 1, 85748 Garching, Germany}
\affiliation{Fakultät für Physik, Ludwig-Maximilians-Universität München, Schellingstraße 4, 80799 Munich, Germany}

\author{Thomas Udem}
\affiliation{Max-Planck-Institut für Quantenoptik, Hans-Kopfermann-Straße 1, 85748 Garching, Germany}
\affiliation{Fakultät für Physik, Ludwig-Maximilians-Universität München, Schellingstraße 4, 80799 Munich, Germany}

\author{Akira Ozawa}
\affiliation{Institute for Multidisciplinary Sciences, Yokohama National University, Yokohama 240-8501, Japan}
\affiliation{Max-Planck-Institut für Quantenoptik, Hans-Kopfermann-Straße 1, 85748 Garching, Germany}

\date{\today}

\renewcommand{\thefigure}{A.\arabic{figure}}
\renewcommand{\thetable}{A.\arabic{table}}
\renewcommand{\theequation}{A.\arabic{equation}}
\setcounter{figure}{0}
\setcounter{table}{0}
\setcounter{equation}{0}

\renewcommand\thesection{\Alph{section}}

\maketitle
\onecolumngrid

\section{Derivation of the Sideband Formula}
\label{ap:derivation}
In the following, we give a detailed derivation of equation~\eqref{eq:cross_section_analytic} of the main text. For this, we first re-derive the analytic expression of the single-ion cross-sections following \cite{winelandLaser1979_sup}. We then generalize the expression to the case of $N$ normal modes, as in the case of an axially excited linear ion chain with $N$ ions.
Consider the dipole transition matrix element between the initial motional state $\ket{n}$ and $\ket{n+\Delta n}$ for a single mode \cite{leibfried_quantum_2003_sup, winelandLaser1979_sup}:

\begin{equation}
    M(n, \Delta n)\coloneq \bra{n+\Delta n}e^{i\eta(\hat{a}+\hat{a}^\dagger)}\ket{n} = e^{-\eta^2/2}(i\eta)^\abs{\Delta n}\sqrt{\frac{n_<!}{n_>!}}L_{n_<}^\abs{\Delta n}(\eta^2)\, ,
\end{equation}
where $n_<$ $(n_>)$ is the smaller (larger) of $n$ and $n+\Delta n$, and $L_n^\alpha(x)$ are the generalized Laguerre-Polynomials. In general, to calculate the cross-section of changing the motional state by $\Delta n$, one must average over the occupation of the allowed initial states $P(n)$:

\begin{equation}
    \label{ap_eq:cross_section_average}
    \sigma(\Delta n) = \sigma_0\sum_{n=n^\prime}^\infty \abs{M(n, \Delta n)}^2 P(n)\, ,
\end{equation}
where $n^\prime = \max(0, -\Delta n)$ and $\sigma_0$ is the cross-section of a single, rigidly fixed ion. We assume $P(n)$ is the thermal distribution given by
\begin{equation}
    \label{ap_eq:thermal_dist}
    P(n) = \left(1-\exp(-\frac{\hbar\omega}{k_B T})\right)\exp(-n\frac{\hbar\omega}{k_B T})\, .
\end{equation}
For $\Delta n\geq0$, we get
\begin{equation}
\begin{aligned}
    \sigma(\Delta n) &= \sigma_0\left(1-\exp(-\frac{\hbar\omega}{k_B T})\right)\sum_{n=0}^\infty\abs{M(n, \Delta n)}^2 \exp(-n\frac{\hbar\omega}{k_B T})\\
    &= \sigma_0 \left(1-\exp(-\frac{\hbar\omega}{k_B T})\right) e^{-\eta^2}\eta^{2\Delta n}
    \sum_{n=0}^\infty \frac{n!}{(n+\Delta n)!} \left(L_n^{\Delta n}(\eta^2)\right)^2 \exp(-n\frac{\hbar\omega}{k_B T})\, .
\end{aligned}
\end{equation}
Following \cite{winelandLaser1979_sup}, this expression can be simplified using the generating function given in \cite[p.~189]{batemanHigher2023_sup}:
\begin{equation}
    \label{eq:gen_function}
    \sum_{n=0}^\infty \frac{n!}{(n+\gamma)!} L_n^\gamma(x) L_n^\gamma(y) z^n = 
    (1-z)^{-1} \exp(-z\frac{x+y}{1-z}) (xyz)^{-\gamma/2} I_\gamma\left(2\frac{(xyz)^{1/2}}{1-z}\right),\,\, \abs{z}<1\, .
\end{equation}
Here $I_\gamma(x)$ is the modified Bessel function of the first kind and order $\gamma$. In our expression we can identify $x=y=\eta^2$, $\gamma=\Delta n$ and $z=\exp(-\frac{\hbar\omega}{k_B T})$. For $\Delta n\geq0$ we then arrive at
\begin{equation}
    \sigma(\Delta n) = \sigma_0 e^{-\eta^2} \exp(-\frac{z}{1-z}2\eta^2) z^{-\Delta n/2} I_{\Delta n}\left(2\eta^2 \frac{z^{1/2}}{1-z}\right)\, .
\end{equation}
In the thermal distribution, the mean motional quantum number is given by
\begin{equation}
\bar{n} = \frac{\exp(-\frac{\hbar\omega}{k_B T})}{1-\exp(-\frac{\hbar\omega}{k_B T})}=\frac{z}{1-z}\, ,
\end{equation}
simplifying the above equation to
\begin{equation}
    \sigma(\Delta n) = \sigma_0 e^{-\eta^2 (1+2\bar{n})} \left(\frac{\bar{n}}{\bar{n}+1}\right)^{-\Delta n/2} I_{\Delta n}\left(2\eta^2 \sqrt{\bar{n}(\bar{n}+1)}\right)\, .
\end{equation}
Analogously, for $\Delta n<0$ we have
\begin{equation}
\begin{aligned}
    \sigma(\Delta n) = \sigma_0 \left(1-\exp(-\frac{\hbar\omega}{k_B T})\right) e^{-\eta^2}\eta^{-2\Delta n}
    \sum_{n=-\Delta n}^\infty \frac{(n+\Delta n)!}{n!} \left(L_{n+\Delta n}^{-\Delta n}(\eta^2)\right)^2 \exp(-n\frac{\hbar\omega}{k_B T})\, .
\end{aligned}
\end{equation}
Note that the sum starts at $n=-\Delta n>0$ as states with $n<0$ do not exist. Shifting the summation index by $\Delta n$, we get 
\begin{equation}
\begin{aligned}
    \sigma(\Delta n) &= \sigma_0 \left(1-\exp(-\frac{\hbar\omega}{k_B T})\right) e^{-\eta^2}\eta^{-2\Delta n} \exp(\Delta n\frac{\hbar\omega}{k_B T})
    \sum_{n=0}^\infty \frac{n!}{(n-\Delta n)!} \left(L_{n}^{-\Delta n}(\eta^2)\right)^2 \exp(-n\frac{\hbar\omega}{k_B T})\\
    &= \sigma_0 e^{-\eta^2(1+2\bar{n})} \left(\frac{\bar{n}}{\bar{n}+1}\right)^{-\Delta n/2} I_{-\Delta n}\left(2\eta^2\sqrt{\bar{n}(\bar{n}+1)}\right)\, .
\end{aligned}
\end{equation}
Since $I_\gamma(x)=I_{-\gamma}(x)$, we have for any integer $\Delta n$
\begin{equation}
\label{ap_eq:cross_sec_single_ion}
\begin{aligned}
    \sigma(\Delta n) = \sigma_0 e^{-\eta^2(1+2\bar{n})} \left(\frac{\bar{n}}{\bar{n}+1}\right)^{-\Delta n/2} I_{\Delta n}\left(2\eta^2\sqrt{\bar{n}(\bar{n}+1)}\right)\, .
\end{aligned}
\end{equation}
Having derived the analytic expression for the spectrum of a single mode, we can now turn to the case of $N$ modes. The first step is to replace the single-mode kick operator with the multi-mode kick operator for the $i$-th ion \cite{herrmannFeasibility2009_sup, morigiTwospecies2001_sup}
\begin{equation}
    e^{i\eta(\hat{a}+\hat{a}^\dagger)} \rightarrow \prod_{\alpha=1}^N e^{i\eta_i^\alpha(\hat{a}_\alpha+\hat{a}_\alpha^\dagger)}\, ,
\end{equation}
with the generalized Lamb-Dicke parameter $\eta_i^\alpha$.
With this, we define
\begin{equation}
\label{ap_eq:matrix_ai}
    M_i^\alpha(n, \Delta n) = \bra{n+\Delta n}e^{i\eta_i^\alpha(\hat{a}_\alpha+\hat{a}_\alpha^\dagger)}\ket{n}
\end{equation}
for the $i$-th ion and mode $\alpha$, and the matrix element
\begin{equation}
\label{ap_eq:matrix_i}
    M_i(\mathbf{n}, \Delta \mathbf{n}) = \bra{\mathbf{n}+\Delta \mathbf{n}}\prod_{\alpha=1}^N e^{i\eta_i^\alpha(\hat{a}_\alpha+\hat{a}_\alpha^\dagger)}\ket{\mathbf{n}}=\prod_{\alpha=1}^N M_i^\alpha(n_\alpha, \Delta n_\alpha)\, ,
\end{equation}
for exciting ($\Delta n>0$) or de-exciting ($\Delta n<0$) the motion of the ion chain from $\ket{\mathbf{n}}$ to $\ket{\mathbf{n}+\Delta\mathbf{n}}$ by electronically exciting ion $i$. Here we defined $\ket{\mathbf{n}}=\ket{n_1, n_2, ..., n_N}$.
The averaged cross section for exciting the $i$-th ion and changing the motional state of the chain by $\Delta \mathbf{n}$ is thus given by
\begin{equation}
\begin{aligned}
\sigma_i(\Delta\mathbf{n}) &= \sigma_0\sum_{\mathbf{n}\in C(\Delta\mathbf{n})} \abs{M_i(\mathbf{n}, \Delta \mathbf{n})}^2 P(\mathbf{n})\, ,
\end{aligned}
\end{equation}
where
\begin{equation}
\begin{aligned}
\label{ap_eq:cart_C}
C(\Delta\mathbf{n})=\prod_{\alpha=1}^N\{n_\alpha\ |\ n_ \alpha \in \mathbb{N}_0\text{ and }n_\alpha\geq n_\alpha^\prime=\max(0, -\Delta n_\alpha)\}
\end{aligned}
\end{equation}
is the Cartesian product of the sets of allowed initial states for $n_1, n_2, \ldots,n_N$ for a given $\Delta\mathbf{n}$. The natural numbers including 0 are denoted by $\mathbb{N}_0$.
Assuming a thermal distribution of the initial states $P(\mathbf{n})=\prod_{\alpha=1}^N P_\alpha(n_\alpha)$ with
\begin{equation}
\begin{aligned}
    P_\alpha(n_\alpha) = \left(1-\exp(-\frac{\hbar\omega_\alpha}{k_B T})\right)\exp(-n_\alpha\frac{\hbar\omega_\alpha}{k_B T})\, ,
\end{aligned}
\end{equation}
we can write
\begin{equation}
\begin{aligned}
\label{ap_eq:cross_section_vec_analytic}
\sigma_i(\Delta\mathbf{n}) &= \sigma_0\!\!\!\!\sum_{\mathbf{n}\in C(\Delta\mathbf{n})} \prod_{\alpha=1}^N \abs{M_i^\alpha(n_\alpha, \Delta n_\alpha)}^2 P_\alpha(n_\alpha)\\
&= \sigma_0 \prod_{\alpha=1}^N\sum_{n_\alpha=n_\alpha^\prime}^\infty \abs{M_i^\alpha(n_\alpha, \Delta n_\alpha)}^2 P_\alpha(n_\alpha)\, .
\end{aligned}
\end{equation}
Every element of the cartesian product $C(\Delta\mathbf{n})$ is reproduced exactly once by multiplying out the second line.
The inner sum on the second line is the same as for the single ion case. Therefore, we use the result from above and find
\begin{equation}
\begin{aligned}
\sigma_i(\Delta\mathbf{n})= \sigma_0 \prod_{\alpha=1}^N e^{-\left(\eta_i^\alpha\right)^2(1+2\bar{n}_\alpha)} \left(\frac{\bar{n}_\alpha}{\bar{n}_\alpha+1}\right)^{-\Delta n_\alpha/2} I_{\Delta n_\alpha}\left(2\left(\eta_i^\alpha\right)^2\sqrt{\bar{n}_\alpha(\bar{n}_\alpha+1)}\right)\, ,
\end{aligned}
\end{equation}
which is equation~\eqref{eq:cross_section_analytic} from the main text.

\renewcommand{\thefigure}{B.\arabic{figure}}
\renewcommand{\thetable}{B.\arabic{table}}
\renewcommand{\theequation}{B.\arabic{equation}}
\renewcommand{\thelstlisting}{B.\arabic{lstlisting}}
\renewcommand{\lstlistingname}{}
\setcounter{figure}{0}
\setcounter{table}{0}
\setcounter{equation}{0}

\section{Proof that the sum over all sidebands is Unity}
\label{ap:proof}
For the $i$-th ion in an ion chain with $N$ modes the sum of all sidebands $\sigma_i(\Delta\mathbf{n})$, enumerated by $\Delta\mathbf{n}\in\mathbb{Z}^N$ with $\mathbb{Z}$ the set of integer numbers, recovers the full line strength $\sigma_0$ of a rigidly fixed ion:
\begin{equation}
\sum_{\Delta\mathbf{n}\in\mathbb{Z}^N}\sigma_i(\Delta\mathbf{n})/\sigma_0=1\, .
\end{equation}
\begin{proof}
Take the definition of the matrix element for the single ion ($N=1$):
\begin{equation}
    M(n, \Delta n)=\bra{n+\Delta n}e^{i\eta(\hat{a}+\hat{a}^\dagger)}\ket{n}\, .
\end{equation}
For any initial state $\ket{n}$, the sum of the squared matrix elements over all sidebands is:
\begin{equation}
    \begin{aligned}
    &\sum_{\Delta n=-n}^\infty \abs{M(n, \Delta n)}^2= \sum_{\Delta n=-n}^\infty \abs{\bra{n+\Delta n}e^{i\eta(\hat{a}+\hat{a}^\dagger)}\ket{n}}^2=\\
    &\sum_{\Delta n=-n}^\infty \bra{n}e^{-i\eta(\hat{a}+\hat{a}^\dagger)}\ket{n+\Delta n}\bra{n+\Delta n}e^{i\eta(\hat{a}+\hat{a}^\dagger)}\ket{n}=\bra{n}e^{-i\eta(\hat{a}+\hat{a}^\dagger)}e^{i\eta(\hat{a}+\hat{a}^\dagger)}\ket{n}\, ,
    \end{aligned}
\end{equation}
where we used $\sum_{n=0}^\infty \ket{n}\bra{n}=\mathbb{1}$. Note that the sum starts at $\Delta n=-n$, as only final states with $n+\Delta n\geq0$ exist. We now use the Baker-Campbell-Hausdorff formula
\begin{equation}
    e^Xe^Y=e^{X+Y+[X,Y]/2}\, ,
\end{equation}
which holds when $[X, [X,Y]]=[Y, [Y, X]]=0$, 
and find
\begin{equation}
    \sum_{\Delta n=-n}^\infty \abs{M(n, \Delta n)}^2= \braket{n}{n} = 1\, .
\end{equation}

For $N$ modes we have for the $i$-th ion, see \eqref{ap_eq:matrix_i}:
\begin{equation}
    M_i(\mathbf{n}, \Delta\mathbf{n})=\prod_{\alpha=1}^N M_i^\alpha(n_\alpha, \Delta n_\alpha)=\prod_{\alpha=1}^N \bra{n_\alpha+\Delta n_\alpha}e^{i\eta^i_\alpha(\hat{a}_\alpha+\hat{a}_\alpha^\dagger)}\ket{n_\alpha}\, .
\end{equation}
For a given initial state $\ket{\mathbf{n}}$, the set of allowed sidebands is given by the cartesian product
\begin{equation}
D(\mathbf{n})=\prod_{\alpha=1}^N\{\Delta n_\alpha\ |\ \Delta n_\alpha \in \mathbb{Z}\text{ and } \Delta n_\alpha\geq-n_\alpha\}\, ,
\end{equation}
ensuring that for the final state $n_\alpha+\Delta n_\alpha\geq0$ for all modes $\alpha$.
The sum of the squared matrix elements over all allowed sidebands for any initial state $\ket{\mathbf{n}}$ is
\begin{equation}
\label{ap_eq:vec_sum_unity}
    \sum_{\Delta \mathbf{n}\in D(\mathbf{n})}\!\!\!\! \abs{M_i(\mathbf{n}, \Delta \mathbf{n})}^2=
    \sum_{\Delta \mathbf{n}\in D(\mathbf{n})}\prod_{\alpha=1}^N\abs{M_i^\alpha(n_\alpha, \Delta n_\alpha)}^2=
    \prod_{\alpha=1}^N\sum_{\Delta n_\alpha=-n_\alpha}^\infty\!\!\!\!\!\abs{M_i^\alpha(n_\alpha, \Delta n_\alpha)}^2 = \prod_{\alpha=1}^N 1 = 1\, .
\end{equation}
The product over the single-mode sums produces every element of the cartesian product $D(\mathbf{n})$ exactly once, in analogy to \eqref{ap_eq:cross_section_vec_analytic}.

The cross-sections are given by
\begin{equation}
\sigma_i(\Delta\mathbf{n})=\sigma_0\!\!\!\!\sum_{\mathbf{n}\in C(\Delta \mathbf{n})}\!\!\!\! \abs{M_i(\mathbf{n}, \Delta\mathbf{n})}^2 P(\mathbf{n})\, ,
\end{equation}
where $P(\mathbf{n})$ is any normalized distribution of the initial states and $C(\Delta \mathbf{n})$ is defined in eq.~\eqref{ap_eq:cart_C}.
The sum over all sidebands is
\begin{equation}
\label{ap_eq:changing_sums_1}
    \sum_{\Delta\mathbf{n}\in \mathbb{Z}^N}\!\!\!\!\sigma_i(\Delta\mathbf{n})/\sigma_0 = \sum_{\Delta\mathbf{n} \in \mathbb{Z}^N}\sum_{\mathbf{n} \in C(\Delta\mathbf{n})}\!\!\!\!\!\abs{M_i(\mathbf{n}, \Delta\mathbf{n})}^2 P(\mathbf{n})\, .
\end{equation}
The constraint $n_\alpha+\Delta n_\alpha\geq 0$ can be imposed on either index: restricting $\mathbf n$ via $C(\Delta\mathbf n)$ or restricting $\Delta\mathbf n$ via $D(\mathbf n)$ enumerates the same pairs $(\mathbf n,\Delta\mathbf n)$, so we may interchange the sums and arrive at
\begin{equation}
     \sum_{\Delta\mathbf{n}\in \mathbb{Z}^N}\!\!\!\!\sigma_i(\Delta\mathbf{n})/\sigma_0 = \sum_{\Delta\mathbf{n} \in \mathbb{Z}^N}\sum_{\mathbf{n} \in C(\Delta\mathbf{n})}\!\!\!\!\!\abs{M_i(\mathbf{n}, \Delta\mathbf{n})}^2 P(\mathbf{n})=\sum_{\mathbf{n} \in \mathbb{N}_0^N} \sum_{\Delta\mathbf{n} \in D(\mathbf{n})}\!\!\!\!\!\abs{M_i(\mathbf{n}, \Delta\mathbf{n})}^2 P(\mathbf{n})=\sum_{\mathbf{n}\in \mathbb{N}_0^N} \!\! P(\mathbf{n}) = 1\, ,
\end{equation}
where we used \eqref{ap_eq:vec_sum_unity} and that $P(\mathbf{n})$ is a normalized distribution.
\end{proof}

\renewcommand{\thefigure}{C.\arabic{figure}}
\renewcommand{\thetable}{C.\arabic{table}}
\renewcommand{\theequation}{C.\arabic{equation}}
\renewcommand{\thelstlisting}{C.\arabic{lstlisting}}
\renewcommand{\lstlistingname}{}
\setcounter{figure}{0}
\setcounter{table}{0}
\setcounter{equation}{0}

\section{Algorithm for the efficient calculation of sidebands}

Outside the Lamb--Dicke regime, the laser excitation couples strongly to the ion motion. Since for $N$ modes, all combinations of sideband transitions can in principle be excited, the number of sidebands that must be taken into account grows exponentially. Even if we limit the occupation-number change per mode to the range $\Delta n=-s_\mathrm{min},\ldots, s_\mathrm{max}$, the total number of sidebands $\Delta \mathbf{n}=(\Delta n_1, \Delta n_2, \ldots \Delta n_N)$ is $(s_\mathrm{min}+s_\mathrm{max}+1)^N$.
As an example, take the evaluation of Fig.~\subfigref[d]{fig:spectrum_panel} of the main text, which shows the spectrum of an ion chain with 15 ions. For the given parameters, we can limit the range to $\Delta n_\alpha=-7,\ldots,9$ to include all sidebands larger than $10^{-6}\sigma_0$. The naive number of sidebands that would need to be evaluated is therefore $17^{15}=2.86\times10^{18}$, which is computationally prohibitive.
However, most of the calculated sidebands contribute only marginally to the spectrum. We implemented an algorithm that efficiently excludes sidebands smaller than a cutoff $\epsilon$. In the example, this reduces the number of sidebands that need to be calculated to $1.2\times10^5$, when the cutoff is set to $\epsilon=10^{-6}$.

The following discussion is limited to the calculation of the single ion cross sections $\sigma_i(\Delta\mathbf{n})$ of the $i$-th ion, using Eq.~\eqref{eq:cross_section_analytic} of the main text. It is straightforward to apply this algorithm to the full ion chain by summing over the single ion contributions, see Eq.~\eqref{eq:total_cross_section} of the main text.

For the algorithm, we exploit that the contributions factorize by mode:
\begin{equation}
\begin{aligned}
\sigma_i(\Delta\mathbf{n})/\sigma_0= \prod_{\alpha=1}^N e^{-\left(\eta_i^\alpha\right)^2(1+2\bar{n}_\alpha)} \left(\frac{\bar{n}_\alpha}{\bar{n}_\alpha+1}\right)^{-\Delta n_\alpha/2} I_{\Delta n_\alpha}\left(2\left(\eta_i^\alpha\right)^2\sqrt{\bar{n}_\alpha(\bar{n}_\alpha+1)}\right) \eqcolon \prod_{\alpha=1}^N K_i^\alpha (\Delta n_\alpha)\, .
\end{aligned}
\end{equation}
For each mode $\alpha$ we first tabulate all $K_i^\alpha(s)$ for $s=-s_\mathrm{min},\ldots, s_\mathrm{max}$, as sketched in Table~\ref{tab:k}. This produces $N(s_\mathrm{min}+s_\mathrm{max}+1)$ entries: For the example, we get $15\times17=255$ entries.
Here it should be noted that by definition $0\leq K_i^\alpha(s)\leq1$, see equation~\eqref{ap_eq:cross_section_vec_analytic}.

\begin{table}[!h]
\caption{Tabulated $K_i^\alpha(s)$ for 15 ions\label{tab:k}}
\begin{tabular}{|l|l|l|l|l|}
 \hline
 $\alpha$\textbackslash $s$ & $-7$ & $-6$ & $\ldots$ & $9$ \\
 \hline
 1 & $K_i^1(-7)$ & $K_i^1(-6)$ & $\ldots$ & $K_i^1(9)$ \\
 \hline
 2 & $K_i^2(-7)$ & $K_i^2(-6)$ & $\ldots$ & $K_i^2(9)$ \\
 \hline
 $\ldots$ & $\ldots$ & $\ldots$ & $\ldots$ & $\ldots$ \\
 \hline
 15 & $K_i^{15}(-7)$ & $K_i^{15}(-6)$ & $\ldots$ & $K_i^{15}(9)$\\
 \hline
\end{tabular}
\end{table} 
From this, sidebands can be constructed as follows: Take any $K_i^1(s)$ from the first line (mode), combine it with any $K_i^2(s)$ from the second line, and so on. This results in the prohibitive number of $17^{15}$ sidebands from the example.
However, while constructing sidebands this way, we can efficiently discard sidebands that are smaller than $\epsilon$: if the partial product over the first $m$ modes is smaller than $\epsilon$, multiplying by any further factors $0\leq K_i^\alpha(s)\leq1$ cannot raise it above $\epsilon$, so all sidebands starting with that combination of $K_i^\alpha(s)$ can be neglected. This procedure allows to reduce the number of sidebands that need to be evaluated drastically: For the example from $17^{15}$ to only $1.2\times10^5$.
\newpage
In detail:
\begin{itemize}
    \item Iterate over $\Delta n_1\in \{-s_\mathrm{min},\ldots, s_\mathrm{max}\}$ and compute $p_1=K_i^1(\Delta n_1)$. If $p_1<\epsilon$, discard all sidebands starting with this $\Delta n_1$. Otherwise continue to mode 2.
    \item For each surviving $\Delta n_1$, iterate over $\Delta n_2$ and compute $p_2=p_1 K_i^2(\Delta n_2)$. If $p_2<\epsilon$, discard all sidebands starting with this $(\Delta n_1, \Delta n_2)$. Otherwise continue to mode 3.
    \item $\ldots$
    \item Continue until all $N$ modes are included. The calculated full products are the sidebands with $p_N\geq\epsilon$.
\end{itemize}

Below we give Python-like pseudocode for the recursive algorithm that returns all sidebands larger than $\epsilon$ for the $i$-th ion. The output is stored in a dictionary \verb|kappa| that maps the occupation-number changes $\Delta\mathbf{n}$ to the normalized sideband strength $\sigma_i(\Delta \mathbf{n})/\sigma_0$. Indices follow this paper’s conventions and therefore do not correspond to valid Python code.

\begin{lstlisting}[caption=Python-like pseudocode., escapeinside={(*}{*)}, language=Python]
# Inputs:
# N                 -- number of modes
# K_i[alpha][s]     -- array of (*$K_i^\alpha(s)$*), with (*$s\in \{-s_\mathrm{min},\ldots,s_\mathrm{max}\}$*)
# s_range           -- list of s values: (*$\{-s_\mathrm{min},\ldots,s_\mathrm{max}\}$*)
# eps               -- cutoff (*$\epsilon$*)

# Output:
# kappa             -- dictionary mapping dn((*$=\Delta\mathbf{n}$*)) to the sideband strength (*$\sigma_i(\Delta \mathbf{n})/\sigma_0$*)

def search(alpha, partial_product, partial_dn, kappa):
    for s in s_range:
        p = partial_product*K_i[alpha][s]
        dn = copy(partial_dn)
        dn.append(s)
        if p >= eps:
            if alpha == N:
                kappa[dn] = p
            else:
                search(alpha+1, p, dn, kappa)
#       else: disregard

# Initialize
kappa = {}
search(1, 1, [], kappa)
\end{lstlisting}
\putbib[apssamp_sup]

\end{bibunit}

\end{document}